# Investigation of level spacing distribution of nuclear energy levels by maximum likelihood estimation method

M. A. Jafarizadeh[a,b *], N. Fouladi[c †], H. Sabri[c], B. Rashidian Maleki[c]

[a] Department of Theoretical Physics and Astrophysics, University of Tabriz, Tabriz 51664, Iran.

[b] Research Institute for Fundamental Sciences, Tabriz 51664, Iran.

[c] Department of Nuclear Physics, University of Tabriz, Tabriz 51664, Iran.

[*] E-mail:jafarizadeh@tabrizu.ac.ir
[†]E-mail:fouladi@tabrizu.ac.ir

# Abstract

The Nearest Neighbor Spacing Distribution (NNSD) is one of the commonly used methods in statistical analysis of regular and chaotic behaviors of nuclear spectra. In this paper, Maximum Likelihood Estimation (MLE) method is proposed to evaluate the parameter of (NNSD)s. We estimate the parameter of (NNSD)s for different mass groups, nuclei with special values of deformation parameter (β) and also for nuclear spectra correspond to three dynamical symmetry limits and transition regions in the framework of the Interacting Boson Model (IBM) (with pure experimental data). The obtained results confirm theoretical predictions even in cases where the small size of data does not allow definite conclusions with the other methods. Also the ML estimated parameters have minimum uncertainty with variations very close (much closer than those obtained by other methods) to the so called Cramer-Rao Lower Bound (CRLB). ML estimated values predict more regular dynamics in compare to what other estimation methods indicate. We have also evaluated CRLB for all distributions (Brody, Berry-Robnik and Abul-Magd) with both values obtained from MLE and in different sequences where Brody distribution has the least CRLB.

PACS: 24.60.Lz, 21.10.Ma, 21.60.Ev

# Introduction

The study of non-linear systems with chaotic behaviors has regarded as one of interesting concepts in recent years. Random Matrix Theory (RMT) has been commonly used for investigating non-fixed properties of very excited nuclei [1-5]. The resonance spacing of scattered neutrons and protons by atomic nuclei, is often introduced as the first application of RMT in nuclear physics, which is first studied by Wigner [1-2]. Despite some developments in experimental applications, the majority of new results from these topics have been investigated by theoretical studies after 1980's. The obtained results from broken symmetries and relationship between chaos and integrable systems allow new applications even from experimental aspects [2]. Though RMT was originally developed for special applications in nuclear physics and then widely applied in other sciences but the small size of experimental data in this branch caused to some unusual uncertainties in the results [3-5].This problem can be considered as the main reason for collecting classes of different nuclei in special mass groups for preparing sequences and studying of level spacing. In RMT, nuclear Hamiltonian is assumed as Gaussian Orthogonal Ensembles (GOE) of random matrices with an anti-unitary symmetry [1]. This form was very successful in describing the systems with time invariance and also exhibited a chaotic dynamics of nuclei with excitation energy near the particle emission threshold [1]. On the other hand, systems whose classical dynamics are everywhere rigorous in the phase space, is well represented by Poisson distribution [1-2]. These classifications in nuclear structure mean, nuclei with definite symmetries can be regarded as integrable systems (regular or Poisson limit) while nuclei with mixed symmetries represent chaotic dynamics [6]. The best model in order to explain these individual and combined symmetries is Interaction Boson Model (IBM) [6-8] which offers an algebraic approach in describing nuclear structure. Also intermediate behavior between these two regular and chaotic limits is usual for different mass groups of nuclei (as represented in [3-5], the lightest nuclei obviously displays chaotic dynamic in comparing to regular behavior of heaviest ones).

Statistical properties, namely the fluctuations of energy levels can be investigated by different methods such as Nearest Neighbor Spacing Distribution (NNSD) [1-5], linear correlation coefficients between adjacent spacing [1-3] ,the Dyson-Mehta $\Delta_3$ statistic [10] and etc. (but we restrict our study to NNSD). Various distribution functions [11-13] with special theoretical aspects were proposed which all of them can describe some behaviors or some special ranges of nuclei [3-5]. The parameters of every distribution can be tuned to interpolate between the Wigner and Poisson distributions. Generally theses parameters are evaluated with least square fitting (LSF) method [3-6], where the great uncertainty in some cases or unacceptable results [3-5], makes almost impossible the lucid conclusions.

On the other hand, LSF is one of widely used estimation methods more well-known than the Maximum Likelihood Estimation (MLE) method or Bayesian Estimation methods (BEM) [13-21], where as it is shown in [20-21], the LSF is really equivalent to producing a maximum likelihood in estimating the variables that are linearly related to some Gaussian case. Therefore, one expects, LSF method yields estimation very close to GOE or Wigner limit, while the other estimation methods, those with more precise estimation in compare to LSF, yield estimation more close to Poison or regular limit.

In some cases where this method couldn't yield appropriate results, other estimation methods such as Maximum Likelihood Estimation (MLE) method or Bayesian Estimation methods (BEM) [13-21] have been applied. In this paper, we have used MLE method in order to evaluate the parameters of all used distributions (Brody, Berry-Robnik and Abul-Magd ones). Also, to obtain the uncertainty of results, we have used Cramer-Rao Lower Bound (CRLB) which is the common method to calculate the uncertainty of unbiased estimators (as mentioned in [19], the MLE method achieves the lowest CRLB).

To feature the advantages of MLE method to other ones, we have compared the results obtained by MLE method for Brody and Berry-Robnik distributions with those investigated in [3-5], where the MLE method yields accuracies very close to CRLB in compare with other methods and predict less chaotic more regular dynamics. Also to make a physical meaning for the results obtained in this paper, we calculate the parameter of Berry-Robnik distribution in sequences constructed of nuclei with special values of deformation parameter,$\beta$, while the resulting values confirm theoretical predictions with again less chaoticity in compare to LSF estimated values.

We have also used the MLE method in estimating parameter of Abul-Magd's distribution in sequences of oblate and prolate nuclei and compared it with [22-23] (which is investigated (BEM)). Also, we have studied nuclei corresponding to three dynamical symmetry limits and nuclei in transitional regions [9,24-26] of IBM (here used sequences constructed from pure experimental data [27-28]), where the estimated values corresponding to the $2^{+}, 4^{+}$ levels, confirm theoretical predictions , namely regular behavior for nuclei with dynamical symmetry (particularly in nuclear with U(5) dynamical symmetry) and also chaotic behavior of nuclei in transitional regions . Finally in order to compare different distributions in the same sequences (prolate, oblate, three dynamical symmetry limits and three transitional regions of IBM ), we have evaluated CRLB for all Brody, Berry-Robnik and Abul-Magd distributions where Brody distribution has the least CRLB.

The paper is arranged as follows. Section II briefly summarizes IBM model (the majority of our results related to this model), unfolding processes (the method of preparing sequences (used data)), well known distributions, MLE method and CRLB. In section III, the results of MLE related to all distribution have been presented. Section IV, contains the numerical results obtained by applying the MLE method to different sequences. Section V is devoted to comparison of MLE method with LSF one, based on results given in section IV. The paper ends with appendices containing the details and related calculations of Brody, Berry-Robnik and Abul-Magd distributions and CRLB.

## II) Preliminaries

This section introduces the notation used in the paper and reviews relevant concepts from Interaction Boson Model (IBM), Nearest Neighbor Spacing Distribution (NNSD), maximum likelihood method and Cramer-Rao Lower Bound (CRLB).

### a) Interaction Boson Model (IBM)

The interacting boson approximation represents a significant step forward in our understanding of nuclear structure. It offers a simple Hamiltonian capable of describing collective nuclear properties across a wide range of nuclei and is founded on algebraic (group) theoretical techniques. The IBM [6-9] is expressed in terms of a U(6) Lie algebra spanned by the bilinear Combinations of five pairs of $L = 2$ ($d$-boson operators) and one pair of $L = 0$ ($s$-boson operators). Thus, the Hilbert space of the model is the carrier space for a so-called symmetric irrep of the compact unitary group U(6). It can be realized as a subspace of the six-dimensional harmonic oscillator Hilbert space. In terms of s- and d-boson operators the most general IBM Hamiltonian can be written as:

$$H = E_0 + c_0 \hat{n}_d + c_2 Q^{\mathcal{X}}.Q^{\mathcal{X}} + c_1 L^2 \ , \tag{1}$$

where $\hat{n}_{\mathrm{d}} = \mathrm{d}^{\dagger}.\tilde{\mathrm{d}}$ is the number operator of d-bosons, L is angular momentum and $Q^{\mathcal{X}}$ is quadrupole operator defined as [7-9]

$$Q^{\mathcal{X}} = \left(d^{\dagger} \times \tilde{s} + s^{\dagger} \times \tilde{d}\right)^2 + \mathcal{X}\left(d^{\dagger} \times \tilde{d}\right)^2, \tag{2}$$

with $\mathcal{X}$ as control parameter. The IBM has three dynamic symmetries corresponding to the following algebra chains

$$U(6) \supset \begin{Bmatrix} U(5) \supset O(5) \\ SU(3) \\ O(6) \supset O(5) \end{Bmatrix} \supset O(3) \qquad \begin{matrix} I \\ II \\ III \end{matrix} \ , \tag{3}$$

Chain (I) occurs for $c_2 = 0$ which describes vibrational nuclei or U(5) limit, chain (II) is obtained for $c_0 = 0$ & $\mathcal{X} = -\frac{\sqrt{7}}{2}$ to describe rotational nuclei or SU(3) limit and chain(III) arises for $\mathcal{X} = 0$ & $c_0 =$ to describe $\gamma$ unstable nuclei or O(6) limit[6-9]. Different nuclei with definite values of ratio $\mathrm{R}_{4/2} = \frac{E(4_1^+)}{\mathrm{E}(2_1^+)}$ can exhibit these three chains and transitional regions which can be obtain from Casten triangle [7-9] in figure 1.

## b) Nearest Neighbor Spacing Distribution (NNSD)

The study of statistical properties of nuclear spectra has long been a subject of great interest and it can be investigated by different methods, such as Nearest Neighbor Spacing Distribution NNSD [1-5], linear correlation coefficients between adjacent spacing [1,3] ,the Dyson-Mehta $\Delta_3$statistic [10]. We focus here on NNSD, which is one of the simplest tools for studying the short-range fluctuations in nuclear spectra. In this method, level spacing of nuclear spectra is compared with RMT predictions. The main point in comparing data to the predictions of RMT is to use some special levels with the same symmetry [3-5] mainly same quantum numbers. This requirement in nuclear physics means to use levels with same total quantum number (J) and same parity, where these group of levels will be called "sequences" [3-5]. Denoting the energy difference between the adjacent levels in every sequence by level distance $S_i$, one can use s=S/D (which D is the average spacing [3] ) to obtain level spacing distribution as a function of dimensionless parameter. For sequences obeying GOE statistics, NNSD probability distribution function is approximated with the Wigner distribution [1]

$$P(s) = \frac{1}{2}\pi s e^{-\frac{\pi s^2}{4}} , \qquad (4)$$

Analysis of the statistical information from proton and neutron resonance for different nuclei demonstrates that the NNSD for levels with excitation energy about 8Mev is well represented by Wigner distribution. On the other hand, for majority of quantum systems with regular behavior in phase space, NNSD can be analyzed with Poisson distribution [1,2]

$$P(s) = e^{-s} , \qquad (5)$$

In most cases, the NNSD distributions show a intermediate behavior between these two limits, namely the Wigner and the Poisson distributions [3-5] (this indicates theoretical predictions about mixture of regular and chaotic dynamics for low-lying energy levels of excited nuclei [3,4]). Different distributions are suggested to investigate intermediate situation of different systems, which have one or more parameters and can exhibit this interpolation between both limits. One of popular distribution is Brody distribution [11]

$$P(s) = b(1+q)s^q e^{-bs^{q+1}} \qquad b = \left[\Gamma\left(\frac{2+q}{1+q}\right)\right]^{q+1} , \qquad (6)$$

that considers a power-law level repulsion and interpolates between the Poisson (q = 0) and Wigner ($q = 1$) distributions .Another distribution is Berry-Robnik distribution which is derived by assuming that the energy level spectrum is a product of the superposition of independent subspectra, which are contributed respectively from localized eigenfunctions onto invariant (disjoint) phase space region[12]

$$P(s) = \left[q + \frac{1}{2}\pi(1-q)s\right] e^{-qs-\frac{1}{4}\pi(1-q)s^2} , \qquad (7)$$

Also another distribution which is appropriate to use, is the NNSD given by the Rosenzweig and Porter random matrix model. The exact form of this model is complicated and its simpler form is proposed by Abul-Magd in Refs.[13.22-23] as:

$$P(s,f) = \left[1 - f + f(0.7 + 0.3f)\frac{\pi s}{2}\right] \times \exp\left(-(1-f)s - f(0.7+0.3f)\frac{\pi s^2}{4}\right), \qquad (8)$$

Where interpolates between Poisson ($f = 0$) and Wigner ($f = 1$) distributions.

## c) Maximum Likelihood Estimation (MLE)

The likelihood function for the probability distribution function $f(x;\theta) = f(x_1, \dots, x_n; \theta)$ of random variables $(X_1, \dots, X_n)$ (both discrete and continuous variables), is defined as [14-21]

$$L(\theta) = f(x;\theta) = f(x_1, \dots, x_n; \theta), \tag{9}$$

That is the chance function for observing variables $(x_1, \dots, x_n)$ in order to obtaining a correct choice of $\theta$. If $\theta_{ML} = s(x_1, \dots, x_n)$ indicates the maximum value of function $L(\theta)$, namely

$$L(\theta_{ML}) = MaxL(\theta), \tag{10}$$

Thus, the Likelihood estimator of $\theta$ is defined as [14-19]

$$\Theta_{ML} = s(X_1, \dots, X_n) \ , \tag{11}$$

Therefore $\theta_{ML}$ is the estimate or MLE suggestion for $\theta$. In evaluating the maximum value of $\theta$, we will consider the fact that $L(\theta)$ and $\ln L(\theta)$have maximum value for the same$\theta$. In the following, we will present some properties of MLE method which can explain the goal of above expressed suggestion [18-21]

- Consistency: probability, the estimator converges to the value that has been estimated
- Asymptotic Normality: with increasing the sample size, the distribution of MLE tends to Gaussian distribution with mean $\theta$ and also covariance matrix becomes equal with the inverse of the Fisher information matrix
- Efficiency: when the sample size tends to infinity, for i.e., it would reach to the lower bund of Cramer-Rao. It means that asymptotic mean squared error of any asymptotically unbiased estimator isn't lower than MLE.

Newton-Raphson iteration method will be used for obtaining the exact result with the minimum variance (for more details see Appendix (C)).

## d) Cramer-Rao Lower Bound (CRLB)

In estimation theory and statistical application, the Cramer–Rao lower bound (CRLB) measures how close this estimator's variance comes to this lower bound. Suppose θ is an unknown deterministic parameter which is to be estimated from measurements of $x$ and also suppose that its corresponding distribution probability density function is $f(x;\theta)$. Inverse of the Fisher information bounds the variance of any unbiased estimator $\hat{\theta}$ of θ as fallow [14-21]:

$$var(\hat{\theta}) \geq \frac{1}{MF(\theta)} \quad , \tag{12}$$

where M is the sample size and Fisher information $F(\theta)$ is defined as fallow:

$$F(\theta) = E\left[\left(\frac{\partial \ln f(x;\theta)}{\theta}\right)^2\right] \quad , \quad f(x;\theta) \equiv P(s) \quad \Longrightarrow \quad F(\theta) = \sum \frac{1}{P(s)}\left[\frac{\mathrm{d} \ln P(s)}{\mathrm{d}\theta}\right]^2,$$

The expression (12) is called the Cramer-Rao inequality. The scalar quantity $\frac{1}{MF(\theta)}$ is the Cramer-Rao lower bound on the variance of unbiased estimators of $f(x;\theta)(\equiv P(s))$ [14].

- **The Cramer-Rao Lower Bound for scalar functions of scalar parameters**

By considering an unbiased estimator $T(X)$ of a function $\psi(\theta)$ of the parameter θ we can obtain a more general form of the bound. Here, by unbiasedness we mean E $\{T(X)\} = \psi(\theta)$. In this case, the bound is given by [14-17]

$$var(\hat{\theta}) \geq \frac{[\psi'(\theta)]^2}{F(\theta)} \quad , \tag{13}$$

where $\psi'(\theta)$ is the derivative of $\psi(\theta)$, and $F(\theta)$ is the Fisher information that was defined above.

- **The Cramer-Rao Lower Bound for vector functions of vector parameters**

Define a parameter column vector in order to extending the Cramér–Rao bound into multiple parameters,

$$\theta = [\theta_1, \theta_2, \dots, \theta_d]^T \in \mathbb{R}^d,$$

with probability density function of $f(x;\theta)$ which satisfies the following regularity condition:

- The Fisher information matrix would be a $d \times d$ matrix with element $F_{m,k}$ that is defined as [14-16]

$$F_{m,k} = E\left[\frac{d}{d\theta_m}\log f(x;\theta)\frac{d}{d\theta_k}\log f(x;\theta)\right],$$

Let $T(X)$ be an estimator of any vector function of parameters, $T(X) = (T_1(X), .., T_N(X))^T$, and denote its expectation vector $E[T(X)]$ by $\rho(\theta)$. The Cramér–Rao bound shows that the covariance matrix $T(X)$ satisfies [14-16]

$$cov_\theta(T(X)) \geq \frac{\partial\rho(\theta)}{\partial\theta^T}[\,F(\theta)]^{-1}\frac{\partial\rho^T(\theta)}{\partial\theta} \quad . \tag{14}$$

While as (12), the expression (14) is called the Cramer-Rao inequality and quantity $\frac{\partial\rho(\theta)}{\partial\theta^T}[\,F(\theta)]^{-1}\frac{\partial\rho^T(\theta)}{\partial\theta}$ is the Cramer-Rao lower bound .

The difference between the traces of left and right sides of equation (14) will be used for presenting the amount of decreasing of the uncertainty variation of estimated parameters during the iterations.

## III) MLE approach to evaluate the parameter of NNSD

Now we are proceeding to determine the parameters of the above introduced nearest neighbor spacing distributions via the maximal likelihood method.

### 1) Brody distribution

Here we propose a generalized Brody distribution with two parameters b and q as [11]:

$$P(s) = b(1+q)s^{q}e^{-bs^{q+1}} \ , \tag{15}$$

Where it reduces to Brody one by choosing $b = \left[\Gamma\left(\frac{2+q}{1+q}\right)\right]^{q+1}$.

Now, in order to estimate the parameters b and q, we need to introduce the corresponding maximum likelihood estimators .To this aim, we try to use the products of the generalized Brody distribution functions as a likelihood function [14-16], namely

$$L(q,b) = \prod_{i=1}^{n} b(1+q)s_i^{q}e^{-bs_i^{q+1}} = [b(1+q)]^n \prod_{i=1}^{n} {s_i}^{q} \ e^{-b\sum s_i^{q+1}}, \tag{16}$$

Then, taking the derivative of the log of likelihood function (16) with respect to the parameters ”q” and ”b” and setting them to zero, i.e., maximizing likelihood function, we obtain the following pair of implicit equations for the required estimators:

$$f_1 \colon \frac{1}{n}\sum s_i^{q+1} - \frac{1}{b} \qquad\qquad for\ b \tag{17}$$

$$f_2 \colon \ \frac{b}{n}\sum \ln s_i \ s_i^{q+1} - \frac{1}{n}\sum \ln s_i - \frac{1}{1+q} \qquad\qquad for\ q \tag{18}$$

Now, the parameters b and q can be estimated by high precision via solving above equation by Newton-Raphson iteration method ( full details are provided in Appedix A), where the difference between the traces of left and right hand sides of equation (14) is used to see the decreasing of the variation of uncertainty of estimated parameters during the iterations.

### 2) Berry-Robnik distribution

We can repeat the above mentioned process for Berry-Robnik distribution [12]

$$P(s) = \left[q + \frac{1}{2}\pi(1-q)s\right] e^{-qs-\frac{1}{4}\pi(1-q)s^2} \tag{19}$$

In order to estimate the parameter of distribution, Likelihood function introduces as product of all P(s) functions [21]

$$L(q) = \prod_{i=1}^{n} P(s_i) = \prod_{i=1}^{n}\left[q + \frac{1}{2}\pi(1-q)s_i\right] e^{-qs_i-\frac{1}{4}\pi(1-q)s_i^2} \tag{20}$$

Then, taking the derivative of the log of likelihood function (20) with respect to its parameter (q) and set it to zero, i.e., maximizing likelihood function, we obtain the following relation for desired estimator (see Appendix (C) for more details)

$$f\colon \sum \frac{1-\frac{1}{2}\pi s_i}{q+\frac{1}{2}\pi(1-q)s_i} - \sum\left(s_i - \frac{1}{4}\pi s_i^2\right) \tag{21}$$

We can estimate q by high accuracy via solving above equation with Newton-Raphson method and also we can use the difference of both sides of equation (13) to obtain the decreasing of uncertainty for estimate values.

### 3) Abul-Magd's distribution

This distribution as Berry-Robnik's one has one parameter [13]

$$P(s,f) = \left[1 - f + f(0.7+0.3f)\frac{\pi s}{2}\right] \times \exp\left(-(1-f)s - f(0.7+0.3f)\frac{\pi s^2}{4}\right)$$

As previous case, we can prepare likelihood function to estimate *f* with product of all $P(s,f)$'s

$$L(f) = \prod_{i=1}^{n} P(s_i) = \prod_{i=1}^{n} \left[1 - f + f(0.7+0.3f)\frac{\pi s_i}{2}\right] e^{-(1-f)s_i - f(0.7+0.3f)\frac{\pi s_i^2}{4}} \tag{22}$$

With setting zero the derivative of the log of likelihood function (22), i.e., maximizing likelihood function, we obtain the following relation for required estimator (see Appendix (C) for more details)

$$estimator\colon \sum \frac{-1 + (0.7+0.6f)\frac{\pi s_i}{2}}{\left[1 - f + f(0.7+0.3f)\frac{\pi s_i}{2}\right]} + \sum s_i - (0.7+0.6f)\frac{\pi s_i^2}{4} \tag{23}$$

We must use Newton-Raphson iteration method to estimate *f* (see Appendix(C) for more details), and also as previous case we can use the difference of right and left sides of (13) to obtain decreasing of uncertainty for our estimates .

## IV) MLE Parameter Estimation of Brody, Berry-Robnik and Abul-Magd distributions from experimental nuclear data and its comparison with other methods.

As mentioned in previous sections, we expect that the estimated values with MLE method yield accuracies which is closer to CRLB. To this aim, we apply the MLE method in estimating the parameter of Brody distribution (A-4,5) by using the data sequences chosen in [3](The most used data sequences in obtaining NNSD of nuclear spectra ). These sequences consist of levels with definite spin and parity of nuclei given in Table1 (this is the copy of Table2 given in Ref.(3)) obtained by applying unfolding processes to the experimental data of Refs. [27,28]. The estimated values of parameter of Brody distribution corresponding to these sequences are listed in tables 1 and 2, respectively, where first is obtained by LSF method while the second one by MLE method. In Maximal likelihood (ML) case we have followed the prescription explained in subsection III.1, namely ML estimated parameters correspond to the converging values of iterations (A-4) and (A-5), where as an initial values we have chosen the values of parameters obtained by LSF method given in Table1, therefore the ML estimated parameters display reduction of uncertainties and yield estimator's variance very close to CRLB as shown in Figures (2,3) respectively for Brody and Berry-Robnik distributions(Due to similarity of CRLB shapes, we only represent two CRLB figures of Tables2,3).

| | **All** | **Even − even** | **Even − even $(0^+, 3^+)$** | **Even − even $(2^+, 4^+)$** | **Even − even not $(2^+, 4^+)$** | **Odd mass** | **Odd − Odd** |
|---|---|---|---|---|---|---|---|
| **All** | $0.43 \pm 0.05$ | $0.42 \pm 0.08$ | $0.56 \pm 0.20$ | $0.34 \pm 0.10$ | $0.56 \pm 0.13$ | $0.40 \pm 0.10$ | $0.44 \pm 0.07$ |
| **Sperical** | $0.60 \pm 0.08$ | $0.55 \pm 0.11$ | $0.52 \pm 0.21$ | $0.52 \pm 0.15$ | $0.57 \pm 0.16$ | $1.06 \pm 0.39$ | $0.60 \pm 0.12$ |
| **Deformed** | $0.30 \pm 0.06$ | $0.26 \pm 0.11$ | $0.74 \pm 0.52$ | $0.16 \pm 0.13$ | $0.51 \pm 0.21$ | $0.32 \pm 0.10$ | $0.31 \pm 0.09$ |
| $\mathbf{0 < A \leq 50}$ | $0.72 \pm 0.16$ | $0.67 \pm 0.25$ | | $0.62 \pm 0.25$ | | | $0.64 \pm 0.21$ |
| $\mathbf{50 < A \leq 100}$ | $0.88 \pm 0.41$ | | | | | | $1.04 \pm 0.67$ |
| $\mathbf{100 < A \leq 150}$ | $0.55 \pm 0.11$ | $0.62 \pm 0.16$ | $0.46 \pm 0.22$ | $0.65 \pm 0.27$ | $0.59 \pm 0.19$ | | $0.47 \pm 0.15$ |
| $\mathbf{150 < A \leq 180}$ | $0.33 \pm 0.07$ | $0.26 \pm 0.11$ | $0.74 \pm 0.52$ | $0.13 \pm 0.14$ | $0.54 \pm 0.22$ | $0.36 \pm 0.14$ | $0.36 \pm 0.11$ |
| $\mathbf{180 < A \leq 210}$ | $0.43 \pm 0.17$ | $0.30 \pm 0.18$ | | $0.16 \pm 0.24$ | | | $1.02 \pm 0.55$ |
| $\mathbf{230 < A}$ | $0.24 \pm 0.11$ | $0.27 \pm 0.32$ | | | | $0.27 \pm 0.16$ | $0.20 \pm 0.16$ |

Table1 [1]: The LSF estimated values of q in Brody distribution for different sequences.

| | **All** | **Even − even** | **Even − even $(0^+, 3^+)$** | **Even − even $(2^+, 4^+)$** | **Even − even not $(2^+, 4^+)$** | **Odd mass** | **Odd − Odd** |
|---|---|---|---|---|---|---|---|
| **All** | $0.101 \pm 0.003$ | $0.029 \pm 0.001$ | $0.214 \pm 0.002$ | $0.078 \pm 0.002$ | $0.104 \pm 0.002$ | $0.179 \pm 0.003$ | $0.214 \pm 0.002$ |
| **Sperical** | $0.315 \pm 0.003$ | $0.281 \pm 0.006$ | $0.224 \pm 0.002$ | $0.218 \pm 0.004$ | $0.263 \pm 0.003$ | $0.477 \pm 0.006$ | $0.304 \pm 0.003$ |
| **Deformed** | $0.194 \pm 0.002$ | $0.141 \pm 0.012$ | $0.371 \pm 0.005$ | $0.199 \pm 0.003$ | $0.207 \pm 0.003$ | $0.130 \pm 0.017$ | $0.308 \pm 0.007$ |
| $\mathbf{0 < A \leq 50}$ | $0.158 \pm 0.001$ | $0.233 \pm 0.002$ | | $0.205 \pm 0.003$ | | | $0.078 \pm 0.002$ |
| $\mathbf{50 < A \leq 100}$ | $0.260 \pm 0.002$ | | | | | | $0.232 \pm 0.005$ |
| $\mathbf{100 < A \leq 150}$ | $0.088 \pm 0.005$ | $0.009 \pm 0.006$ | $0.235 \pm 0.002$ | $0.015 \pm 0.008$ | $0.062 \pm 0.003$ | | $0.177 \pm 0.002$ |
| $\mathbf{150 < A \leq 180}$ | $0.118 \pm 0.003$ | $0.088 \pm 0.002$ | $0.085 \pm 0.005$ | $0.205 \pm 0.002$ | $0.149 \pm 0.003$ | $0.215 \pm 0.002$ | $0.348 \pm 0.003$ |
| $\mathbf{180 < A \leq 210}$ | $0.51 \pm 0.002$ | $0.79 \pm 0.014$ | | $0.223 \pm 0.002$ | | | $0.414 \pm 0.002$ |
| $\mathbf{230 < A}$ | $0.054 \pm 0.002$ | $0.176 \pm 0.002$ | | | | $0.044 \pm 0.003$ | $0.212 \pm 0.003$ |

Table [2]: The ML estimated values of q in Brody distribution for the same sequences used in table 1. Here every cell of table exhibits q-parameter of Brody distribution in every sequence. All sequences are collected from different nuclei with the method introduced in [3].

In Berry-Robnik distribution case, using the sequences of Ref. [5], the corresponding ML iterations given in (C-5) yields the estimated parameters listed in Table (3). Again we have chosen the parameters estimated by LSF method of Ref. (5) as initial values for the iterations.

| Nuclei | q(obtained from fit) | q (obtained from MLE) |
|---|---|---|
| **A** $< 50$ | $0.03 \pm 0.16$ | $0.32 \pm 0.08$ |
| **50** $<$ **A** $< 100$ | $0.27 \pm 0.30$ | $0.69 \pm 0.06$ |
| **100** $<$ **A** $< 150$ | $0.37 \pm 0.32$ | $0.77 \pm 0.10$ |
| **150** $<$ **A** $< 180$ | $0.53 \pm 0.10$ | $0.91 \pm 0.04$ |
| **180** $<$ **A** $< 210$ | $0.27 \pm 0.27$ | $0.82 \pm 0.08$ |
| **230** $<$ **A** | $0.59 \pm 0.18$ | $0.95 \pm 0.19$ |
| **Deformed** $(0^+, 3^+)$ | $0.29 \pm 0.09$ | $0.84 \pm 0.10$ |
| **Spherical** $(2^+, 4^+)$ | $0.34 \pm 0.20$ | $0.63 \pm 0.05$ |
| **Deformed** $(2^+, 4^+)$ | $0.74 \pm 0.23$ | $0.56 \pm 0.21$ |

Table [3]: Comparison of the ML estimated values of q in Berry-Robnik distribution with those estimated by LSF method for different sequences given in Ref. [4,5].

Considering the estimated q values (by MLE and LSF methods) given in tables 1,2 and 3, we can deduce the following important facts:

I) Estimated values of parameters of both Brody and Berry-Robnik distributions given in Tables 1,2 and 3 (estimated by LSF and ML methods, respectively), imply that the distribution of lightest nuclei ($A \leq 50$) displays chaotic behavior in compare to heaviest ones ($230 < A$). But the ML estimated values of parameter in both distributions are less than LSF estimated ones for behavior nuclei, consequently the spectrum of heaviest nuclei are more regular than what LSF estimation indicates. Similarly, in light nuclei, ML estimated values of parameter in both distributions is less than LSF estimated ones, hence the spectrum of light nuclei are not so much chaotic that LSF estimation indicates.

II) As it is predicted in Ref. [3]( by LSF estimation method), $2^+$and $4^+$levels are more regular (are more closer to Poisson distribution) than $0^+$and $3^+$ones. Obviously ML estimated parameters given in Table 2 confirm this but again all of above levels are more regular than what LSF indicates.

III) The nuclei in $150 < A < 180$ sequence (spherical nuclei) are located between two sequence of deformed nuclei, namely $150 < A < 180$ and $> 230$ . As it is shown in Ref. [3, 24 ]( by LSF estimation method), there is a considerable variation in the values of q, that is its values drop to half value in the two neighboring mass region which implies that, spherical nuclei exhibit more chaotic dynamics in compare to deformed ones. The ML estimated values given in Table 2 confirm this behavior but predict less chaotic dynamic for all sequence.

IV) Similarly, the spherical odd-mass and odd-odd nuclei display more chaotic dynamics in compare to deformed nuclei, where ML estimated values given in tables 2 and 3 confirm this but predict less chaotic dynamic again.

V) As already mentioned in previous sections, due to presence of noticeable uncertainty in LSF estimated values (because of High level variance of estimators), it is almost impossible to do any reliable statistical analysis of odd-odd nuclei in mass regions $150 < A < 180$ and $180 < A < 210$, while as a result of small variation in MLE method, hence minimum uncertainty, the trustworthy analysis is quite possible (See table 2 for these reliable ML estimated q values for above mass region). On the whole, ML estimated values are almost exact in all sequences, even in cases with small sample sizes, where by LSF estimation method one cannot achieve the appropriate results.

Also, we can compared ML and LSF estimated q values for different groups of nuclei with a given deformation parameter. As it is predicted by LSF method in Ref. [5], chaoticity degrees of sequence decrease with increasing of β. The same behavior can be seen in ML estimated q values of Berry-Robnik distribution given in Table 4, but with less chaoticity degrees. Again these values are obtained by using the iterations given in (C-5) by choosing LSF estimated q values as an initial values.

| **Different Nuclei** | **q** | $\langle\boldsymbol{\beta}\rangle$ |
|---|---|---|
| $\mathbf{A} < 50$ | $0.32 \pm 0.08$ | **-0.025** |
| $\mathbf{50} < \mathbf{A} < 100$ | $0.69 \pm 0.06$ | **0.032** |
| $\mathbf{100} < \mathbf{A} < 150$ | $0.77 \pm 0.10$ | **0.051** |
| $\mathbf{150} < \mathbf{A} < 180$ | $0.91 \pm 0.04$ | **0.246** |
| $\mathbf{180} < \mathbf{A} < 210$ | $0.82 \pm 0.08$ | **-0.125** |
| $\mathbf{230} < \mathbf{A}$ | $0.95 \pm 0.19$ | **0.217** |

Table [4] :The ML estimated values of q in Berry-Robnik distribution for different sequences (mass groups) of nuclei with given values of β . These sequences are prepared by the same method introduced in Ref.[5].

Also, in Abul-Magd distribution case, we use the sequences of oblate and prolate nuclei introduced in [22]. These are the sequences used in Ref. [22] for estimating the parameter ***f*** of Abul-Magd distribution by Bayesian estimation method (BEM). The corresponding ML iterations given in (C-11) yields the estimated ***f*** values listed in Table 5 and the corresponding NNSD distributions displayed in Fig. 4. As for the initial values of iterations, we have chosen both LSF and BEM (the values given in Ref. [22]) estimated values of parameter ***f***, where both choice yield almost the same values given in Table 5.

| **Nuclei** | $\boldsymbol{f}$(**obtained from fit**) | $\boldsymbol{f}$(**obtained from BEM**) | $\boldsymbol{f}$(**obtained from MLE**) |
|---|---|---|---|
| **Prolate Nuclei** | $0.78 \pm 0.09$ | $0.73 \pm 0.05$ | $0.64 \pm 0.02$ |
| **Oblate Nuclei** | $0.61 \pm 0.09$ | $0.59 \pm 0.07$ | $0.57 \pm 0.07$ |

Table [5]: ML , LSF and BEM estimated values of **f** in Abul-Magd distribution for sequences of oblate and prolate nuclei given in Ref. [22], where BEM estimated values are those of Ref. [22] .

The above given ML estimated values together with NNSD distributions displayed in Fig 4, like the previous distributions, reveals some regularity in oblate sequence in compare to prolate one, but similar to other cases with less chaoticity in both sequences. Also, the ML estimated parameters have the least amount of uncertainties since the variance of estimator is very close to CRLB as shown in Figures 5 (Due to similarity of CRLB shapes, we only represent CRLB figure for oblate nuclei). where the minimum CRLB correspond to for the final value in iteration procedure (ML estimated value), while BEM estimated values correspond to initial values of iteration with variances far from the CRLB. Therefore, one can conclude that, MLE method yield the most exact result in compare to BEM and LSF estimation methods and display less chaoticity in compare to those methods.

Also, in order to investigate level spacing distribution of different nuclei in IBM framework, we have chosen the sequences of Nuclei with specified symmetries listed in Table6. These nuclei correspond to well-known three dynamical symmetry limits and transitional regions of IBM characterized with $R_{4/2} = \frac{E(4_1^+)}{E(2_1^+)}$ ratio. To this aim, following the method introduced in [3-5], we have prepared sequences with $2^+, 4^+$ levels from the experimental data [27-28] and then have estimated q parameter in Brody distribution by using the iterations (A-4,5) of corresponding estimators obtained from MLE method (see Fig. 7). Analogous to the theoretical anticipations [9,25-26], the NNSD distributions of these six regions displayed in Fig. 7 and also estimated values listed in Table 6, indicate that nuclei with U(5) dynamical symmetry [24-26] have maximum regularity in compare to other dynamical symmetry limits[6-9,24-26]. Again MLE method shows more regularity in above three symmetric limits. On the other hand, nuclei in transitional regions have chaotic behavior in compare to dynamical symmetry limits [9, 25-26]. Again ML yield estimator's variance very close to CRLB as shown in Figure 8 for SU(3) symmetry limit and $U(5) - O(6)$ transition regions.

| Special group | Nuclei |
|---|---|
| U(5) | $^{98}$Mo, $^{100}$Mo, $^{108}$Cd, $^{112}$Cd, $^{114}$Cd, $^{110}$Cd, $^{116}$Cd, $^{118}$Cd, $^{118}$Te, $^{120}$Te, $^{122}$Te, $^{124}$Te, $^{126}$Te, $^{112}$Sn, $^{114}$Sn, $^{134}$Xe, $^{154}$Dy,… |
| O(6) | $^{56}$Fe, $^{78}$Ge, $^{80}$Se, $^{130}$Ba, $^{132}$Ba, $^{132}$Ce, $^{134}$Ce, $^{196}$Hg, $^{194}$Pt, $^{196}$Pt, $^{198}$Pt, $^{198}$Hg,… |
| SU(3) | $^{166}$Er, $^{176}$Hf, $^{180}$W, $^{168}$Yb, $^{174}$Hf, $^{160}$Dy, $^{230}$Th, $^{184}$W, $^{232}$Th, $^{182}$W, $^{232}$U, $^{178}$Hf, $^{170}$Yb, $^{162}$Dy, $^{234}$U, $^{164}$Dy, $^{172}$Yb, $^{240}$Pu, $^{168}$Er, $^{170}$Er, $^{246}$Cm,… |
| SU(3)-U(5) | Nd-Sm-Gd isotopes |
| U(5)-O(6) | Ru-Pd isotopes, Xe isotopes(else ones mentioned in above series), $^{134}$Ba |
| O(6)-SU(3) | Os-Pt isotopes(else $^{194}$Pt, $^{196}$Pt, $^{198}$Pt) |

Table [6]: The sequences of Nuclei with specified dynamical symmetries.

| Nuclei | q (obtained from fit) | q (obtained by MLE) |
|---|---|---|
| **Nuclei with O(6) symmetry** | $0.52 \pm 0.16$ | $0.48 \pm 0.06$ |
| **Nuclei with SU(3)symmetry** | $0.71 \pm 0.13$ | $0.57 \pm 0.08$ |
| **Nuclei with U(5) symmetry** | $0.46 \pm 0.18$ | $0.33 \pm 0.06$ |
| **Nuclei with U(5) − O(6) transition** | $0.78 \pm 0.13$ | $0.64 \pm 0.09$ |
| **Nuclei with U(5) − SU(3) transition** | $0.74 \pm 0.14$ | $0.63 \pm 0.10$ |
| **Nuclei with O(6) − SU(3) transition** | $0.94 \pm 0.14$ | $0.74 \pm 0.08$ |

Table [7]: Values of q for sequences introduced in Table6, estimated by LSF and ML methods, respectively.

ML estimated values listed in Tables 4,5 and 7, more or less confirm the above deduced facts from the contents of Tables 1 and 2, namely, due to existence of noticeable uncertainty in LSF estimated values, reliable statistical analysis would be somehow impossible. On the other hand, as a result of small variation in MLE method, hence minimum uncertainty, the trustworthy analysis is possible. In general, ML estimated values are almost exact in all sequences, even in cases where one cannot reach the appropriate results by LSF or BEM estimation methods and also ML estimated values indicate less chaotic dynamics in compare to what LSF or BEM indicates.

In order to compare different distributions, we evaluate Cramer-Rao lower bound, namely the term defined on the right hand side of Cramer-Rao inequality (12) as

$$CRLB \equiv \frac{1}{MF(\theta)}\bigg|_{for\ final\ value\ of\ "\theta" obtained from\ MLE\ or\ fitting\ processes}$$

For all distributions utilized here in this paper. To this aim, we have evaluated the Cramer-Rao lower bound for the Brody, Berry-Robnik and Abul-Magd distributions based on the sequences of prolate, oblate, three dynamical symmetry and also three transitional regions of IBM, (please see Appendices B and C for details ) as tabulated below in Tables 8, 9. The above obtained result implies that, the MLE method yields good accuracies in all distributions, since the ML estimated values have the least uncertainty, that is, the variances in MLE method are more closer to the Cramer-Rao lower bound than those of LSF method as we were expecting. Also, Brody distribution has the least CRLB in compare to two other distributions; hence one can conclude that, it is the best NNSD distribution based on the existing theoretical and experimental data.

| Nuclei | Brody distribution | Berry-Robnik distribution | Abul-Magd distribution |
|---|---|---|---|
| **Oblate nuclei** | q= 0.73 | q=0.37 | f=0.61 |
| | CRLB = $1.38 \times 10^{-10}$ | CRLB =$4.1 \times 10^{-5}$ | CRLB=$1.5 \times 10^{-3}$ |
| **Prolate nuclei** | q=0.82 | q=0.21 | f=0.78 |
| | CRLB =$4.3 \times 10^{-22}$ | CRLB =$1.8 \times 10^{-5}$ | CRLB=$4.5 \times 10^{-4}$ |
| **Nuclei with U(5) symmetry** | q=0.46 | q=0.44 | f=0.57 |
| | CRLB =$2.5 \times 10^{-6}$ | CRLB =$5.5 \times 10^{-5}$ | CRLB=$4.7 \times 10^{-3}$ |
| **Nuclei with SU(3) symmetry** | q=0.71 | q=0.31 | f=0.69 |
| | CRLB =$1.1 \times 10^{-10}$ | CRLB =$3.1 \times 10^{-5}$ | CRLB=$8.1 \times 10^{-4}$ |
| **Nuclei with O(6) symmetry** | q=0.52 | q=0.40 | f=0.62 |
| | CRLB =$5.2 \times 10^{-7}$ | CRLB =$6.8 \times 10^{-5}$ | CRLB=$4.2 \times 10^{-3}$ |
| **Nuclei in U(5)-O(6) region** | q=0.78 | q=0.22 | f=0.80 |
| | CRLB =$1.8 \times 10^{-12}$ | CRLB =$8.4 \times 10^{-5}$ | CRLB=$2.7 \times 10^{-3}$ |
| **Nuclei in U(5)-SU(3) region** | q=0.74 | q=0.29 | f=0.71 |
| | CRLB =$7.04 \times 10^{-14}$ | CRLB =$6.1 \times 10^{-5}$ | CRLB=$1.3 \times 10^{-3}$ |
| **Nuclei in SU(3)-O(6) region** | q=0.94 | q=0.15 | f=0.86 |
| | CRLB =$7.1 \times 10^{-21}$ | CRLB =$1.1 \times 10^{-4}$ | CRLB=$2.5 \times 10^{-3}$ |

Table [8]: CRLBs for LSF estimated parameters in different distributions.

| Nuclei | | Brody distribution | Berry-Robnik distribution | Abul-Magd distribution |
|---|---|---|---|---|
| **Oblate nuclei** | | q= 0.59 | q=0.44 | f=0.57 |
| | | CRLB = $2.5\times10^{-23}$ | CRLB =$1.8\times10^{-5}$ | CRLB=$1.2\times10^{-3}$ |
| **Prolate nuclei** | | q=0.72 | q=0.36 | f=0.64 |
| | | CRLB =$1.2\times10^{-39}$ | CRLB =$1.4\times10^{-5}$ | CRLB=$2.1\times10^{-4}$ |
| **Nuclei with U(5) symmetry** | | q=0.33 | q=0.64 | f=0.35 |
| | | CRLB =$1.2\times10^{-6}$ | CRLB =$4.5\times10^{-5}$ | CRLB=$1.5\times10^{-3}$ |
| **Nuclei with SU(3) symmetry** | | q=0.57 | q=0.46 | f=0.56 |
| | | CRLB =$2\times10^{-21}$ | CRLB =$2.4\times10^{-5}$ | CRLB=$4.5\times10^{-4}$ |
| **Nuclei with O(6) symmetry** | | q=0.48 | q=0.57 | f=0.44 |
| | | CRLB =$3.5\times10^{-8}$ | CRLB =$5.7\times10^{-5}$ | CRLB=$1.6\times10^{-3}$ |
| **Nuclei in U(5)-O(6) region** | | q=0.64 | q=0.34 | f=0.67 |
| | | CRLB =$1.1\times10^{-25}$ | CRLB =$7.5\times10^{-5}$ | CRLB=$1.4\times10^{-3}$ |
| **Nuclei in U(5)-SU(3) region** | | q=0.63 | q=0.40 | f=0.61 |
| | | CRLB =$8.3\times10^{-25}$ | CRLB =$4.9\times10^{-5}$ | CRLB=$7.5\times10^{-4}$ |
| **Nuclei in SU(3)-O(6) region** | | q=0.74 | q=0.30 | f=0.71 |
| | | CRLB =$6.5\times10^{-74}$ | CRLB =$9.6\times10^{-5}$ | CRLB=$1.1\times10^{-3}$ |

Table [9]: CRLBs for ML estimated parameters in different distributions.

One of the widely used method for comparing different distributions is Kullback-Leibler Divergence or Kullback-Leibler Distance (KLD). KLD is a non-symmetric measuring method of distance between the difference two probability distributions $P$ and $Q$. For two probability distributions $P$ and $Q$ with discrete random variables, KLD is defined as (see Appendix(C) for more details)

$$D_{KL}(P\|Q)=\sum_i P(i)\log\frac{P(i)}{Q(i)} \quad , \tag{24}$$

All distribution used in this paper, namely Brody, Berry-Robnik and Abul-Magd distributions are one-parameter distributions and as their parameter varies, they intermediate between Poisson and Wigner distributions, namely the distribution correspond to the limiting values of parameter. Fortunately, in these distribution, KLD reveals the same behavior as the parameter does. Namely, for the values of their parameter near one of the limits (estimated by all three estimation methods MLE, BEM and LSF), KLD displays closer distances to the corresponding limiting distributions. As we pointed out already in previous sections, ML estimated values exhibit more regularity in all sequences, therefore, one can conclude that ML estimated values have closer Kullback-Leibler Distance to Poisson limit.

## V. Conclusions

In the present paper, MLE method is utilized in investigating the statistical properties of nuclear spectra in NNSD framework. Using MLE method, we have estimated the parameter of all used distributions (Brody, Berry-Robnik and Abul-Magd) in sequences of different mass groups, nuclei with special values of deformation parameter (β), oblate and prolate nuclei and also for nuclei correspond to three dynamical symmetry limits and transition regions in the framework of the Interacting Boson Model (IBM) (with pure experimental data).

In all cases, the ML estimated values have minimum uncertainty in compare to those estimated by other methods, that is, the variation of ML estimated values are rather small and close enough to CRLB. Therefore, in investigating the statistical properties of nuclear spectra, MLE method is more reliable than other estimation methods, particularly LSF one, such that ML estimated values indicate more regular dynamics in compare to what LSF or BEM indicates. This is more obvious in cases with small size of data, such that LSF estimated values are not reliable at all. Finally, besides reliability, MLE method is also more handy than other sophisticated estimation methods such as BEM.

# Appendix

## Appendix A

Brody distribution

As mentioned in previous sections, our used distribution has some differences in compare to certain distributions. This is because of troubles which occur in Likelihood function but in the following we will display, our choice doesn't have any difference with the main one

$$P(s) = b(1+q)s^q e^{-bs^{q+1}} \qquad (A-1)$$

$$L(q,b) = \prod_{i=1}^{n} b(1+q)s_i^q e^{-bs_i^{q+1}} = [b(1+q)]^n \left(\prod_{i=1}^{n} s_i{}^q\right) e^{-b\sum s_i^{q+1}} \qquad (A-2)$$

$$\frac{\partial \ln \mathrm{L(q,b)}}{\partial b} = 0 \Rightarrow f_1 : \frac{1}{n}\sum s_i^{q+1} - \frac{1}{b}$$

$$\frac{\partial \ln \mathrm{L(q,b)}}{\partial q} = 0 \Rightarrow f_2 : \frac{b}{n}\sum \ln s_i\, s_i^{q+1} - \frac{1}{n}\sum \ln s_i - \frac{1}{1+q}$$

And we can get the final result by using of Newton-Raphson iteration method:

$$\begin{bmatrix} \mathrm{q_{new}} \\ \mathrm{b_{new}} \end{bmatrix} = \begin{bmatrix} \mathrm{q_{old}} \\ \mathrm{b_{old}} \end{bmatrix} - \mathrm{Df}^{-1}(q_{old}, b_{old}) f(q_{old}, b_{old}) \qquad (A-3)$$

$$\mathrm{Df}(q_{old}, b_{old}) = \begin{bmatrix} \frac{\partial f_1(q_{old}, b_{old})}{\partial q} & \frac{\partial f_1(q_{old}, b_{old})}{\partial b} \\ \frac{\partial f_2(q_{old}, b_{old})}{\partial q} & \frac{\partial f_2(q_{old}, b_{old})}{\partial b} \end{bmatrix} = \begin{bmatrix} \frac{1}{n}\sum \ln s_i\, s_i^{q+1} & \frac{1}{b^2} \\ \frac{b}{n}\sum (\ln s_i)^2\, s_i^{q+1} + \frac{1}{(1+q)^2} & \frac{1}{n}\sum \ln s_i\, s_i^{q+1} \end{bmatrix}$$

$$\mathrm{q_{new}} = q_{old} - \frac{\left[\frac{1}{n}\sum \ln s_i s_i^{q_{old}+1}\right]\left[\frac{1}{n}\sum s_i^{q_{old}+1} - \frac{1}{b_{old}}\right] - \frac{1}{b_{old}{}^2}\left[\frac{b_{old}}{n}\sum \ln s_i s_i^{q_{old}+1} - \frac{1}{n}\sum \ln s_i - \frac{1}{1+q_{old}}\right]}{\left[\frac{1}{n}\sum \ln s_i s_i^{q_{old}+1}\right]^2 - \frac{1}{b_{old}{}^2}\left[\frac{b_{old}}{n}\sum (\ln s_i)^2 s_i^{q_{old}+1} + \frac{1}{(1+q_{old})^2}\right]} \qquad (A-4)$$

$$\mathrm{b_{new}} = b_{old} - \qquad (A-5)$$

$$-\frac{\left[-\frac{b_{old}}{n}\sum (\ln s_i)^2\, s_i^{q_{old}+1} - \frac{1}{(1+q_{old})^2}\right]\left[\frac{1}{n}\sum s_i^{q_{old}+1} - \frac{1}{b_{old}}\right] + \left[\frac{1}{n}\sum \ln s_i\, s_i^{q_{old}+1}\right]\left[\frac{b_{old}}{n}\sum \ln s_i\, s_i^{q_{old}+1} - \frac{1}{n}\sum \ln s_i - \frac{1}{1+q_{old}}\right]}{\left[\frac{1}{n}\sum \ln s_i\, s_i^{q_{old}+1}\right]^2 - \frac{1}{b_{old}{}^2}\left[\frac{b_{old}}{n}\sum (\ln s_i)^2\, s_i^{q_{old}+1} + \frac{1}{(1+q_{old})^2}\right]}$$

And if we plot (b)' values in all iteration stages as have displayed in Figure (9), obviously it would have the same behavior as coefficient of main distribution which we have changed it in our calculations.

## AppendixB

CRLB's calculation

$$cov_\theta\big(T(X)\big) \geq \frac{\partial\rho(\theta)}{\partial\theta^T}\,[\,F(\theta)]^{-1}\,\frac{\partial\rho^T(\theta)}{\partial\theta} \qquad (B-1)$$

$$\theta_1 \to b, \theta_2 \to q$$

$$\rho_1 \to \frac{1}{b} \Rightarrow \quad \frac{\partial\rho_1}{\partial b} = \frac{-1}{b^2}\ , \quad \frac{\partial\rho_1}{\partial q} = 0 \ \ \& \ \ \rho_2 \to \frac{1}{1+q} \Rightarrow \quad \frac{\partial\rho_2}{\partial b} = 0\ , \quad \frac{\partial\rho_2}{\partial q} = \frac{-1}{(1+q)^2}$$

And for Fisher information

$$F(\theta) = \begin{bmatrix} E\left[\left(X_q - \bar{X}_q\right)^2\right] & E[(X_q - \bar{X}_q)(X_b - \bar{X}_b)] \\ E[(X_q - \bar{X}_q)(X_b - \bar{X}_b)] & E[(X_b - \bar{X}_b)^2] \end{bmatrix} \qquad (B-2)$$

In the above relation have

$$X_b = \frac{\partial \ln L(q,b)}{\partial b} = \frac{n}{b} - \sum s_i^{q+1} \qquad \& \quad \bar{X}_b = \frac{1}{n}\sum X_b$$

$$X_q = \frac{\partial \ln L(q,b)}{\partial q} = \frac{n}{1+q} + \sum \ln s_i - b\sum \ln s_i\, s_i^{q+1} \qquad \& \qquad \bar{X}_q = \frac{1}{n}\sum X_q \qquad (B-3)$$

And our used estimator functions for minimum variation are

$$f_1: q - \left(s_i^{q+1} - \frac{1}{b}\right)\ \& \ f_2: b - \left(b\ \ln s_i\, s_i^{q+1} - \ \ln s_i - \frac{1}{1+q}\right) \qquad (B-4)$$

$$cov_\theta\big(T(X)\big) = \begin{bmatrix} E\left[\left(f_1 - \bar{f}_1\right)^2\right] & E[(f_1 - \bar{f}_1)(f_2 - \bar{f}_2)] \\ E[(f_1 - \bar{f}_1)(f_2` - \bar{f}_2)] & E\left[\left(f_2 - \bar{f}_2\right)^2\right] \end{bmatrix}$$

And with proposed functions and derivatives, we have

$$\begin{bmatrix} E\left[\left(f_1 - \bar{f}_1\right)^2\right] & E[(f_1 - \bar{f}_1)(f_2 - \bar{f}_2)] \\ E[(f_1 - \bar{f}_1)(f_2 - \bar{f}_2)] & E\left[\left(f_2 - \bar{f}_2\right)^2\right] \end{bmatrix} \geq \qquad (B-5)$$

$$\frac{1}{E\left[\left(X_q - \bar{X}_q\right)^2\right]\ E[(X_b - \bar{X}_b)^2] - (E[(X_q - \bar{X}_q)(X_b - \bar{X}_b)])^2} \times$$

$$\times \begin{bmatrix} \dfrac{E[(X_b - \bar{X}_b)^2]}{b^4} & -\dfrac{E[(X_q - \bar{X}_q)(X_b - \bar{X}_b)]}{b^2(1+q)^2} \\ -\dfrac{E[(X_q - \bar{X}_q)(X_b - \bar{X}_b)]}{b^2(1+q)^2} & \dfrac{E\left[\left(X_q - \bar{X}_q\right)^2\right]}{(1+q)^4} \end{bmatrix}$$

**Appendix C**

I) In this section as same as App (A) and App(B) ,we replay our calculation for Berry-Robnik distribution

$$P(s) = \left[q + \frac{1}{2}\pi(1-q)s\right] e^{-qs-\frac{1}{4}\pi(1-q)s^2} \qquad (C-1)$$

$$L(q) = \prod_{i=1}^{n} P(s_i) = \prod_{i=1}^{n} \left[q + \frac{1}{2}\pi(1-q)s_i\right] e^{-qs_i-\frac{1}{4}\pi(1-q)s_i^2} \qquad (C-2)$$

$$\ln L(q) = \sum_{i=1}^{n} \ln\left[q + \frac{1}{2}\pi(1-q)s_i\right] - \sum_{i=1}^{n} qs_i + \frac{1}{4}\pi(1-q)s_i^2$$

$$\frac{\mathrm{d}\ln L(q)}{\mathrm{d}q} = \sum \frac{1-\frac{1}{2}\pi s_i}{q + \frac{1}{2}\pi(1-q)s_i} - \sum\left(s_i - \frac{1}{4}\pi s_i^2\right) \qquad \rightarrow F(q) \qquad (C-3)$$

And with Newton-Raphson iteration method, we can get final result as

$$q_{new} = q_{old} - \frac{F(q_{old})}{F'(q_{old})} \qquad (C-4)$$

$$q_{new} = q_{old} - \frac{\sum \frac{1-\frac{1}{2}\pi s_i}{q_{old} + \frac{1}{2}\pi(1-q_{old})s_i} - \sum s_i + \frac{1}{4}\pi s_i^2}{\sum \frac{-\left(1-\frac{1}{2}\pi s_i\right)^2}{\left(q_{old} + \frac{1}{2}\pi(1-q_{old})s_i\right)^2}} \qquad (C-5)$$

And CRLB for Berry-Robnik distribution

$$Var(\hat{\theta}) \geq \frac{1}{MF(\theta)} \qquad \theta \rightarrow q \qquad (C-6)$$

$$F(\theta) = \sum \frac{1}{P(s)}\left[\frac{\mathrm{d}\ln P(s)}{\mathrm{d}\theta}\right]^2 \qquad \& \qquad M = number\ of\ sample \qquad (C-7)$$

$$f_1 = \frac{1-\frac{1}{2}\pi s_i}{q + \frac{1}{2}\pi(1-q)s_i}$$

$$Var(\hat{\theta}) = \frac{1}{n}\sum\left(f_1 - \bar{f}_1\right)^2$$

**II) Abul-Magd's distribution [10,17]**

$$P(s,f) = \left[1 - f + f(0.7 + 0.3f)\frac{\pi s}{2}\right] \times \exp\left(-(1-f)s - f(0.7 + 0.3f)\frac{\pi s^2}{4}\right)$$

$$L(f) = \prod_{i=1}^{n} P(s_i) = \prod_{i=1}^{n} \left[1 - f + f(0.7 + 0.3f)\frac{\pi s_i}{2}\right] e^{-(1-f)s_i - f(0.7+0.3f)\frac{\pi s_i^2}{4}} \qquad (C-8)$$

$$\ln L(f) = \sum_{i=1}^{n} \ln\left[1 - f + f(0.7 + 0.3f)\frac{\pi s_i}{2}\right] - \sum_{i=1}^{n} (1-f)s_i + f(0.7 + 0.3f)\frac{\pi s_i^2}{4} \qquad (C-9)$$

$$\frac{\mathrm{d}\ln L(f)}{\mathrm{d}f} = \sum \frac{-1 + (0.7 + 0.6f)\frac{\pi s_i}{2}}{\left[1 - f + f(0.7 + 0.3f)\frac{\pi s_i}{2}\right]} + \sum s_i - (0.7 + 0.6f)\frac{\pi s_i^2}{4} \quad \rightarrow F(f) \qquad (C-10)$$

$$f_{\text{new}} = f_{old} - \frac{F(f_{old})}{F'(f_{old})}$$

$$= f_{old} - \frac{\sum \frac{-1 + (0.7 + 0.6f_{old})\frac{\pi s_i}{2}}{[1 - f_{old} + f_{old}(0.7 + 0.3f_{old})\frac{\pi s_i}{2}]} + \sum s_i - (0.7 + 0.6f_{old})\frac{\pi s_i^2}{4}}{\sum \frac{[0.3\pi s_i]\left[1 - f_{old} + f_{old}(0.7 + 0.3f_{old})\frac{\pi s_i}{2}\right] - [-1 + (0.7 + 0.6f_{old})\frac{\pi s_i}{2}]^2}{[1 - f_{old} + f_{old}(0.7 + 0.3f_{old})\frac{\pi s_i}{2}]^2} - \sum 0.15\pi s_i^2} \qquad (C-11)$$

As explained for Berry- Robnik distribution, we can introduce CRLB for Abul-Magd's distribution

$$Var(\hat{\theta}) \geq \frac{1}{MF(\theta)} \qquad \theta \rightarrow f$$

$$F(\theta) = \sum \frac{1}{P(s)}\left[\frac{\mathrm{d}\ln P(s)}{\mathrm{d}\theta}\right]^2 \quad \& \quad M = number\ of\ sample \qquad (C-12)$$

$$f_1 = \frac{-1 + (0.7 + 0.6f)\frac{\pi s_i}{2}}{\left[1 - f + f(0.7 + 0.3f)\frac{\pi s_i}{2}\right]} + s_i - (0.7 + 0.6f)\frac{\pi s_i^2}{4}$$

$$Var(\hat{\theta}) = \frac{1}{n}\sum\left(f_1 - \bar{f}_1\right)^2$$

III) Newton-Raphson method

If we have n non-linear equations from n variables

$$f_i(x_1, \dots, x_n) = 0 \qquad for\ i = 1 \ \ to \ \ n$$

Now if $x \in R^n$ and $f: R^n \rightarrow R^n$, we have

$$f(x) = \begin{bmatrix} f_1(x_1, \dots, x_n) \\ \vdots \\ f_n(x_1, \dots, x_n) \end{bmatrix} \qquad (C-13)$$

And finally, we can introduce Jacobian matrices

$$Df(x) = \begin{bmatrix} \frac{\partial f_1}{\partial x_1} & \cdots & \frac{\partial f_1}{\partial x_n} \\ \vdots & \vdots & \vdots \\ \frac{\partial f_n}{\partial x_1} & \cdots & \frac{\partial f_n}{\partial x_n} \end{bmatrix}$$

Now if $f: R^n \to R^n$ is a derivability function in $\hat{x} \in R^n$ if we expand $x$ about $\hat{x}$, we can obtain in matrices form

$$f(x) \approx f(\hat{x}) + Df(\hat{x})(x - \hat{x}) \qquad (C-14)$$

**D) Kullback-Leibler Divergence (KLD)**

The K-L divergence of the probability distributions P and Q of a discrete random variable is defined as:

$$D_{KL}(P\|Q) = \sum_i P(i) \log \frac{P(i)}{Q(i)} \qquad (C-15)$$

In a word, in the average of the logarithmic difference between the probabilities *P* and *Q*, the average is taken by using of the probabilities *P*. The K-L divergence only can be defined if the sum of *P* and *Q* be 1 and if $Q(i) > 0$ for any $i$ , such that $P(i) > 0$. If the 0log0 is appeared in the formula, it would be interpreted as zero.

For distributions *P* and *Q* of a continous random variable, KL-divergence is defined to be an integral :

$$D_{KL}(P\|Q) = \int_{-\infty}^{+\infty} p(x) \log \frac{p(x)}{q(x)} \, dx \qquad (C-16)$$

Where $p$ and $q$ denote the densities of *P* and *Q*.

More generally, if *P* and *Q* are probability measures over a set *X*, and *Q* is absolutely continuous with respect to *P*, then the Kullback–Leibler divergence from *P* to *Q* is defined as

$$D_{KL}(P\|Q) = -\int_x \log \frac{dQ}{dP} dP$$

where $\frac{dQ}{dP}$ is the Radon-Nikodym derivative of *Q* with respect to *P*, and it provides an expression that is exist in the right-hand of the equation. Likewise, if *P* is absolutely continuous with respect to *Q*, then

$$D_{KL}(P\|Q) = \int_x \log\frac{dP}{dQ}\,dP = \int_x \frac{dP}{dQ}\,\log\frac{dP}{dQ}\,dQ \qquad (C-17)$$

We can introduce some properties of KLD as:

- The Kullback–Leibler divergence is always non-negative,

$$D_{KL}(P\|Q) \geq 0$$

a result known as Gibbs' inequality, and $D_{KL}(P\|Q)$ would be zero if and only if $P = Q$.

- The Kullback–Leibler divergence remains well-defined for continuous distributions, and furthermore, under the parameter transformations it would be invariant

- The Kullback–Leibler divergence is additive for independent distributions in a very similar way as Shannon entropy. If $P_1$,$P_2$ are independent distributions, with the joint distribution $P(x,y) = P_1(x)P_2(y)$, and $Q$,$Q_1$,$Q_2$ likewise, then

$$D_{KL}(P\|Q) = D_{KL}(P_1\|Q_1) + D_{KL}(P_2\|Q_2)$$

## Figure caption

**Figure1**. Casten triangle [14], three vertices show dynamical symmetry limits and three edges show transitional regions, respectively**.**

**Figure2.** (Color online)**.** Plots of CRLB for q estimated values of Brody distribution for spherical and deformed nuclei listed in table2. In both graphs, the horizontal axis represents number of iteration and vertical one, represents $Tr[(Cov(T(X))^2] - Tr[(\frac{\partial\rho}{\partial\theta^T} F_\theta^{-1} \frac{\partial\rho^T}{\partial\theta})^2]$.

**Figure3**. (Color online). Plots of CRLB for q estimated values of Berry-Robnik distribution for two different mass regions ( $A \leq 50$ $and$ $A \geq 230$) listed table3. In both graphs, the horizontal axis represents number of iteration and vertical one represents $[ Var(\hat{\theta})] - [\frac{1}{MF(\theta)}]$.

**Figure4**. (Color online). Plots of NNSDs for sequences of oblate and prolate nuclei given in Ref. [10]

**Figure5**. (Color online). Plot of CRLB for ***f*** estimated values of Abul-Magd distribution for oblate nuclei listed in Table 5, where the horizontal axis represents number of iteration and vertical one represents$[ Var(\hat{\theta})] - [\frac{1}{MF(\theta)}]$.

**Figure6.** (Color online). Plots of NNSD for different nuclei correspond to three Limits (dynamical symmetries) of IBM and transitional regions between these limits (all sequences are prepared by experimental data given in Refs.[27,28])

**Figure7.** (Color online). Plots of CRLB for q estimated distribution for nuclei with SU(3) symmetry and also for nuclei in transitional region between two U(5) and O(6) limits**.** In both graphs, the horizontal axis represents number of iteration and vertical one, represents$Tr[(Cov(T(X))^2] - Tr[(\frac{\partial\rho}{\partial\theta^T} F_\theta^{-1} \frac{\partial\rho^T}{\partial\theta})^2]$.

**Figure8**. (Color online). Variation of our proposed constant for Brody distribution in different iteration stages which verify our aim that any change does not occur with the main distribution. The horizontal axis represents number of iteration and vertical one represents $\frac{b}{\left[\Gamma\left(\frac{2+q}{1+q}\right)\right]^{q+1}}$.

**Figure1:**

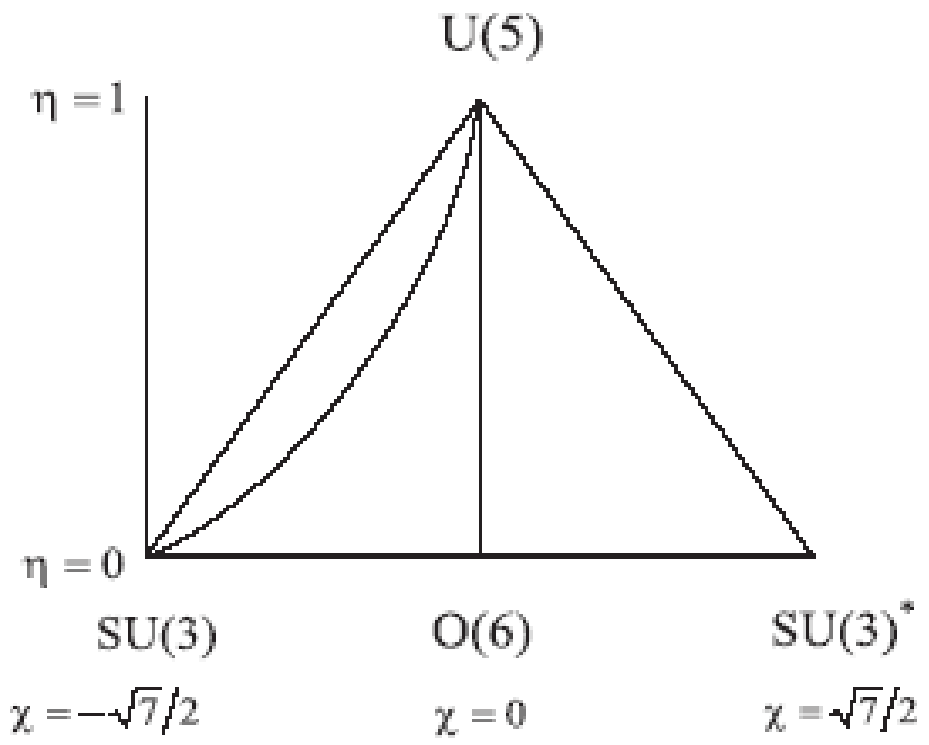

**Figure2**

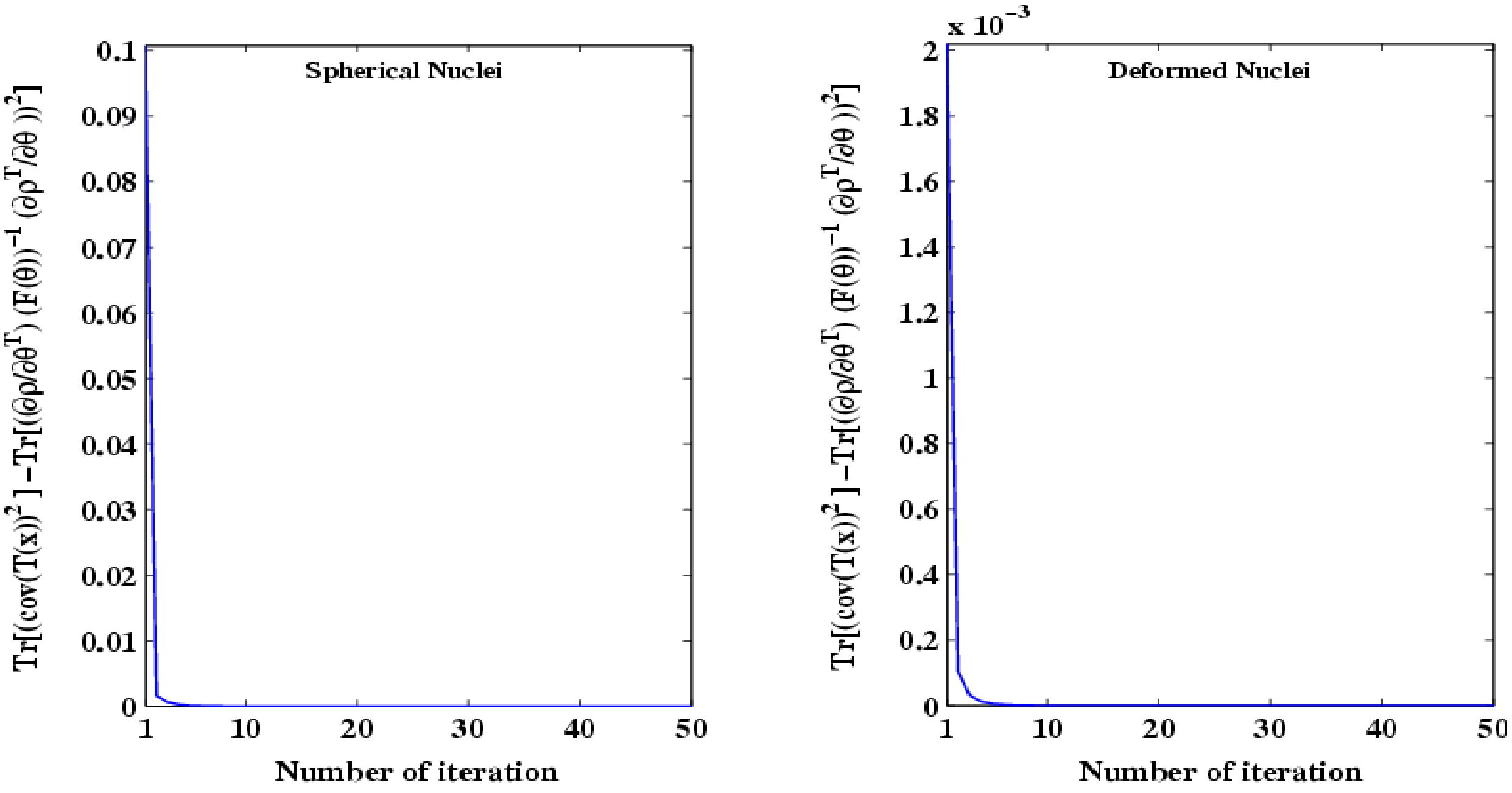

**Figure3**

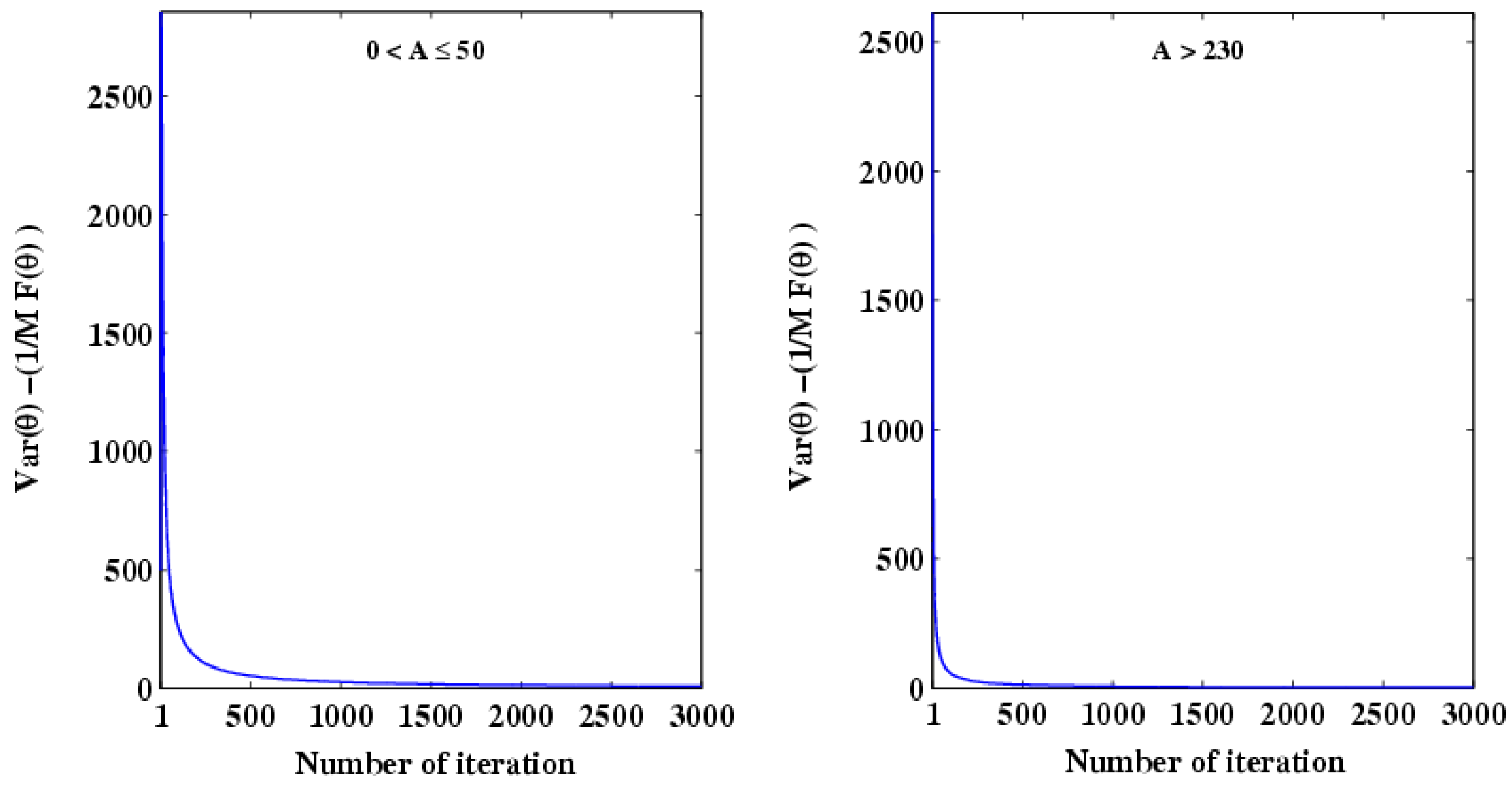

**Figure4**

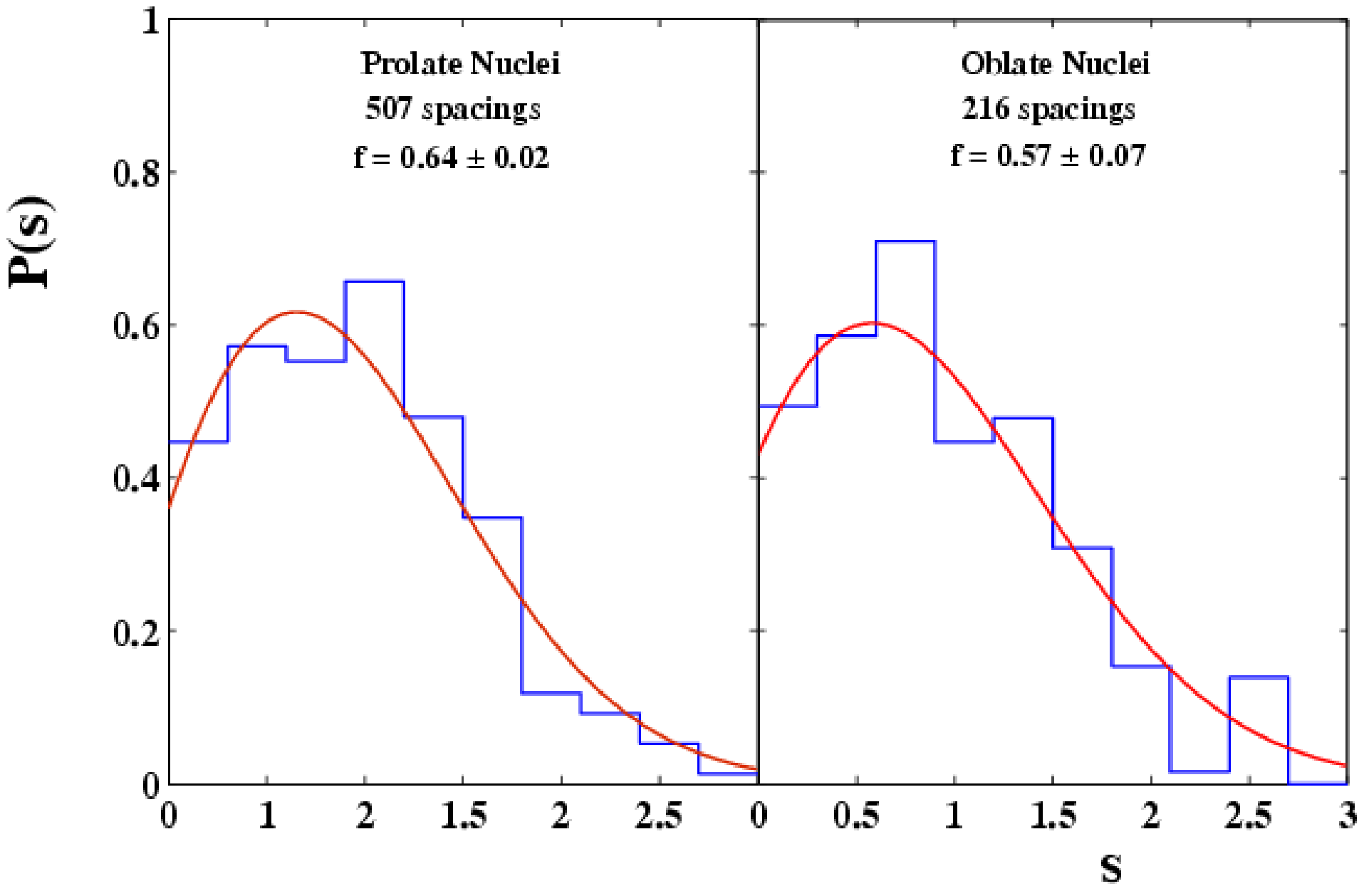

**Figure5**

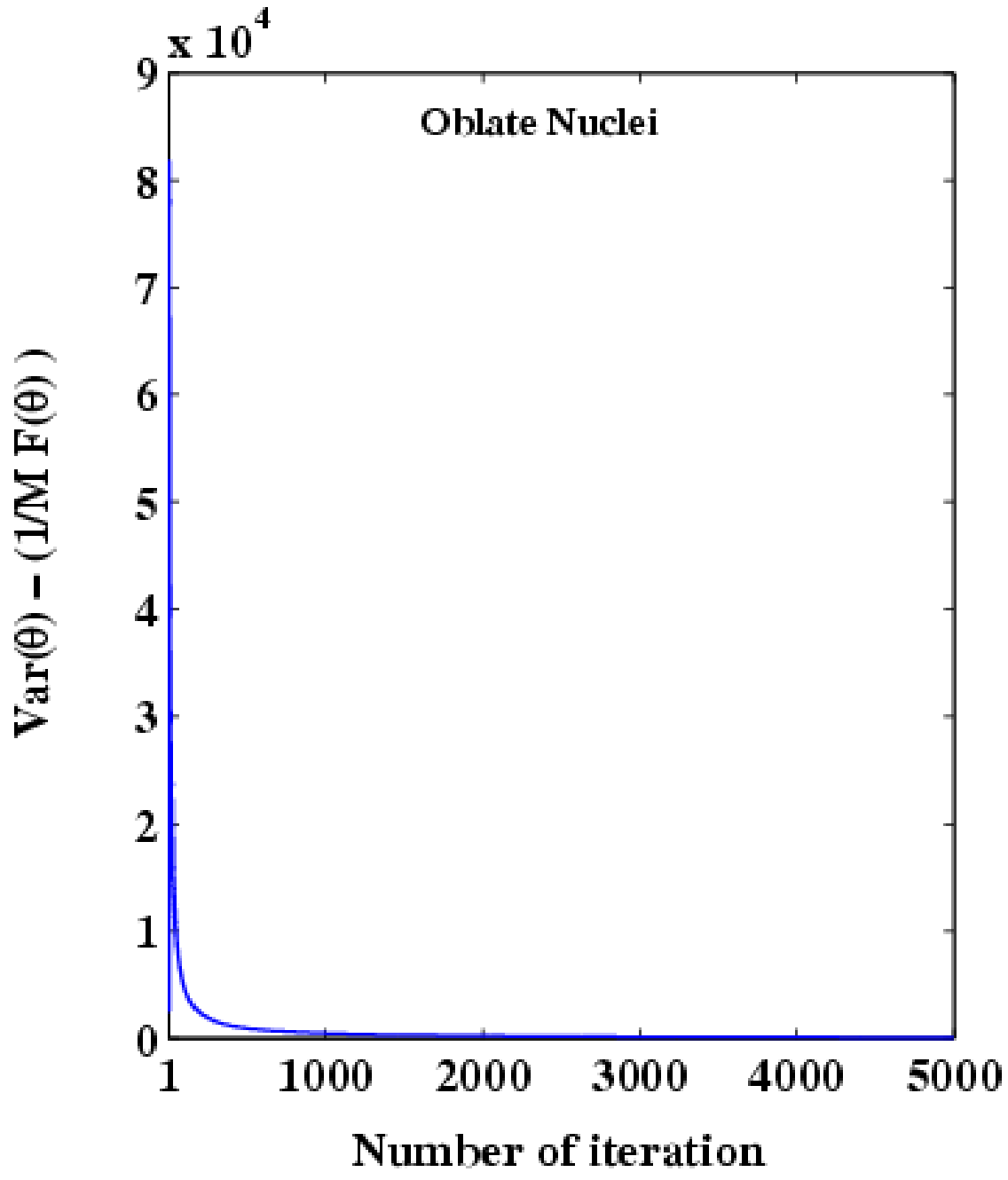

**Figure6**

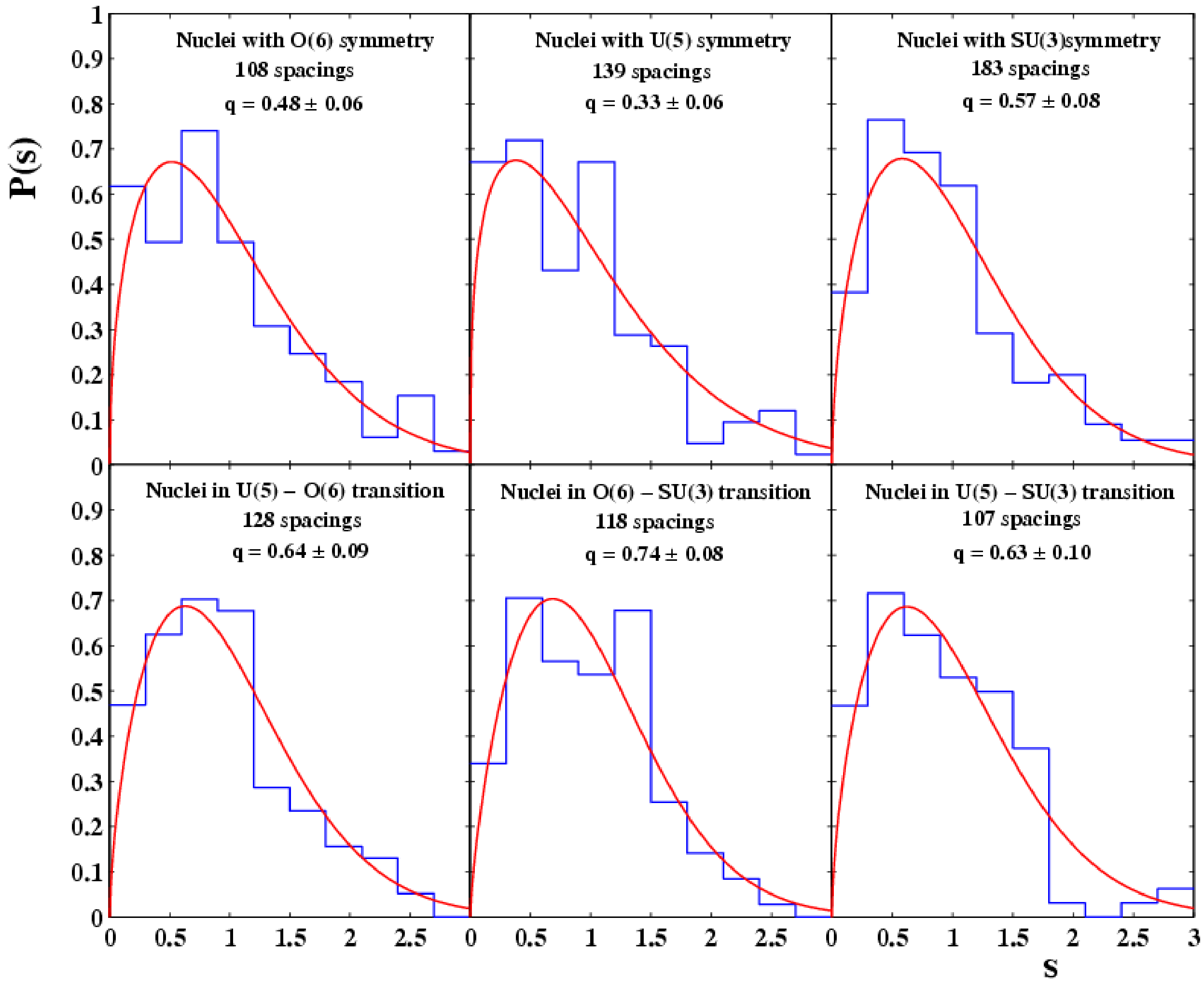

Nuclei with O(6) symmetry
108 spacings
q = 0.48 ± 0.06
Nuclei with U(5) symmetry
139 spacings
q = 0.33 ± 0.06
Nuclei with SU(3)symmetry
183 spacings
q = 0.57 ± 0.08
Nuclei in U(5) – O(6) transition
128 spacings
q = 0.64 ± 0.09
Nuclei in O(6) – SU(3) transition
118 spacings
q = 0.74 ± 0.08
Nuclei in U(5) – SU(3) transition
107 spacings
q = 0.63 ± 0.10
P(s)
S
0
0.1
0.2
0.3
0.4
0.5
0.6
0.7
0.8
0.9
1
0.5
1.5
2
2.5
3

**Figure7**

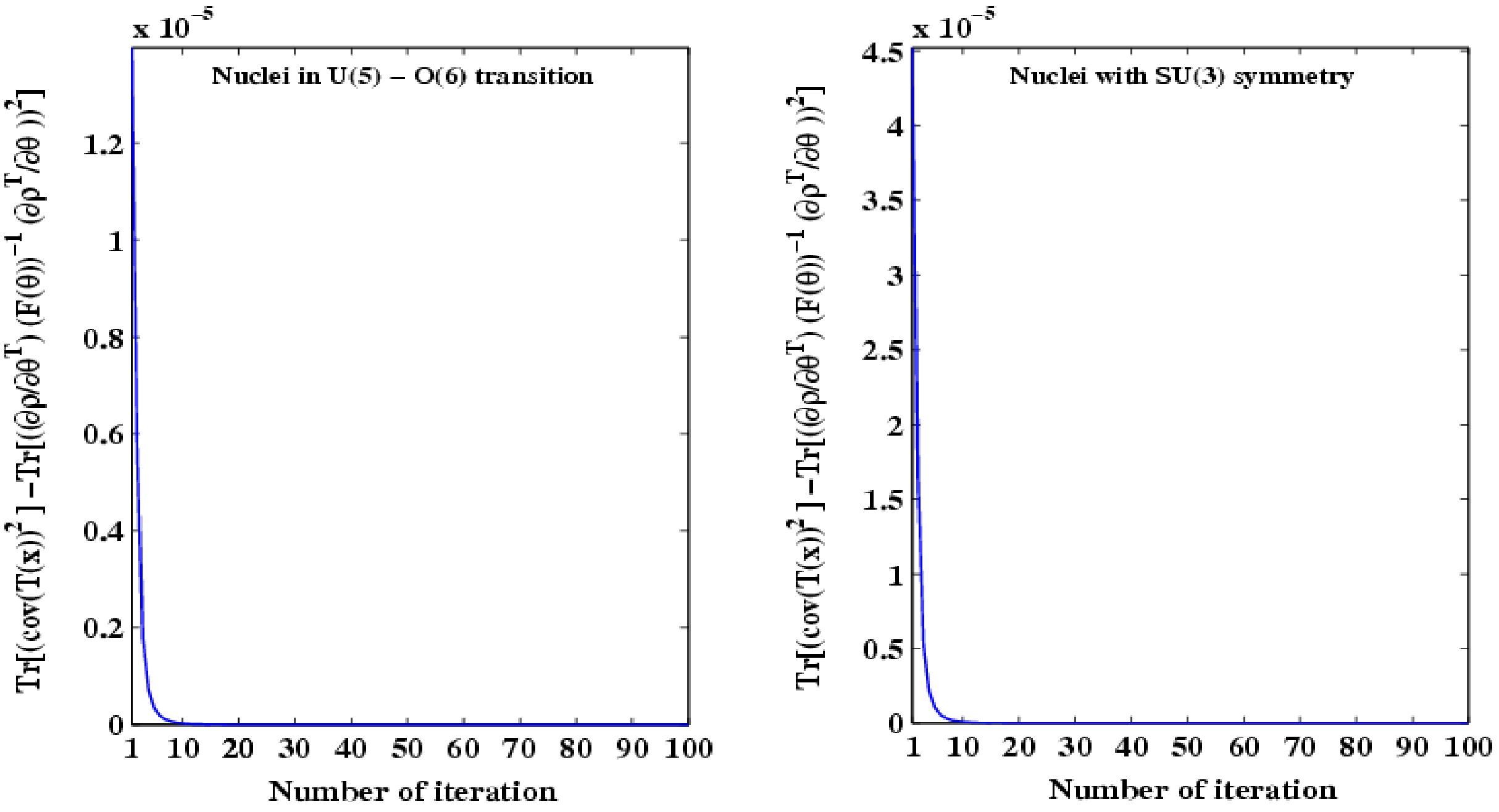

**Figure8**

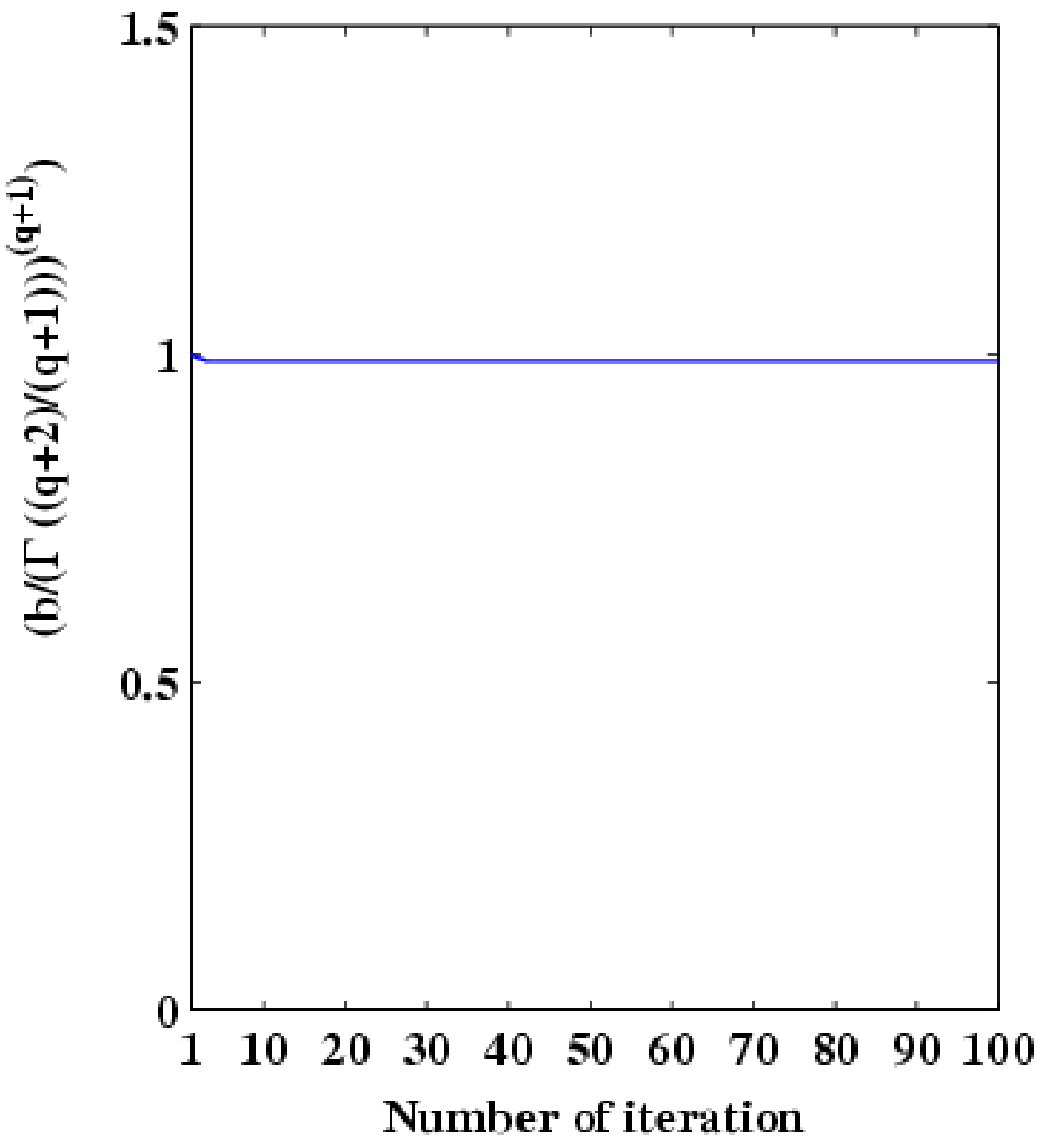